\documentclass[12pt]{article}
\usepackage[english]{babel}
\usepackage{amssymb,amsmath,natbib}
\usepackage[hang,small,tight]{subfigure}
\usepackage{lscape}
\usepackage{multirow}
\usepackage{multirow}
\usepackage{lscape}
\usepackage{graphicx}
\usepackage{hhline}
\usepackage{color}
\def\dir{./}

\newcommand{\bfalpha}{\mbox{\boldmath $\alpha$}}

\newcommand{\bfTheta}{\mbox{\boldmath $\Theta$}}
\newcommand{\bfbeta}{\mbox{\boldmath $\beta$}}

\newcommand{\bfdelta}{\mbox{\boldmath $\delta$}}

\newcommand{\bfgamma}{\mbox{\boldmath $\gamma$}}

\newcommand{\bfomega}{\mbox{\boldmath $\omega$}}

\title{A  Hierarchical Dynamic Beta Regression Model of \\  School Performance in the Brazilian Mathematical Olympiads for Public Schools}
\author{Alexandra M. Schmidt$^{1}$\footnote{{{\it Address for correspondence}: Alexandra M.
Schmidt, Departamento de M\'etodos Estat\'{\i}sticos, Universidade
Federal do Rio de Janeiro, Caixa Postal 68530, Rio de Janeiro, RJ,
Brazil. CEP 21945-970. {\it E-mail}: {\tt
alex@im.ufrj.br}. {\it Homepage}: www.dme.ufrj.br/$\sim$alex}},  Caroline P. de Moraes, and Helio S. Migon \\
\textit{Universidade Federal do Rio de Janeiro, Brazil}}

\date{July 2015}

\begin{document}

\maketitle

\begin{abstract}
The Brazilian  Mathematical Olympiads for Public Schools (OBMEP) is held every year since  2005. In the 2013 edition  there were over 47,000 schools registered  involving nearly 19.2 million students. The Brazilian public educational system is structured into three administrative levels: federal, state and municipal. Students participating in the OBMEP come from three educational levels, two in primary and one in secondary school. We aim at studying the performance of Brazilian public schools which have been taking part of the OBMEP from 2006 until 2013. We propose a standardization of the mean scores of schools per year and educational level which is modeled through a hierarchical dynamic beta regression model. Both the mean and precision  of the beta distribution are modeled as a function of covariates whose effects evolve smoothly with time. Results show that, regardless of the educational level, federal schools have better performance than municipal or state schools. The mean performance of schools increases with the human development index (HDI) of the municipality the school is located in. Moreover, the difference in mean performance between federal and state or municipal schools tends to increase with the HDI. Schools with higher proportion of boys tend to have better mean performance in the second and third educational levels of OBMEP. 

\end{abstract}

\section{Introduction \label{sec:motivation}}

The International Mathematical Olympiad (IMO) is organized yearly (except for 1980) since 1959, when it was first held in Romania. Participation in an IMO is by invitation only. Country's contestants should be selected through that Country's national Mathematical Olympiad or an equivalent selection programme. Brazil has been participating in the IMO since 1979, when it organized the first Brazilian Mathematical Olympiad (OBM).  OBM is organized by the Brazilian Mathematical Society (SBM).  The aims of OBM are to stimulate the study of mathematics by students, develop and improve the training of teachers, influence the improvement of education, in addition to discovering young talents. Brazil has took part in 35 editions of the IMO, since then it won 9 gold, 33 silver, 68 bronze medals, and 29 honourable mentions. The position of Brazil in the unofficial mark ordering available from the IMO website ({\tt https://www.imo-official.org/results.aspx}) shows that it has been performing reasonably well, especially when one considers that Brazil is a developing country that faces many challenges in its educational system.
 
Among the 65 economies  that participated in  the 2012 edition of the Program for International Student Assessment (PISA),   Brazil performed below the average in mathematics, occupying the 58$^{th}$ position in the mathematics mean score. 
On the other hand, in this same edition of PISA, the Brazilian mean mathematics performance reached 391 score points, having increased 35 points compared to the 2003 edition. The 2012 PISA report points out that between 2003 and 2012, performance gains in Brazil are largely attributed to a reduction in the proportion of low-performing (those who perform below the baseline Level 2) students. Indicators from 2012 of the Organisation for Economic Co-operation and Development (OECD) show that Brazil boasts one of the largest increase in expenditure on education between 2000 and 2009 among the countries for which they had data available. Although Brazil's spending on education as a percentage of GDP is below the OECD average, there has been a steady increase in the percentage of GDP invested in education, particularly between 2000 and 2009. Brazil increased public spending on education from 10.5\% of total public expenditure in 2000, to 14.5\% in  2005, and to 16.8\% in 2009 - one of the steepest rates of growth among the 33 countries for which data were available. Brazil ranks $4^{th}$ in this measure out of the 32 countries for which data are  available and above the OECD average of 13\% \citep{OECD:2012}. One of such  investments in education is the Mathematical Olympiads for Public Schools ({\em Olimp\'iada Brasileira de Matem\'atica em Escolas P\'ublicas} or OBMEP). 

\subsection{The Brazilian Mathematical Olympiads for Public Schools}

OBMEP has been promoted since 2005 by the Ministries of Science and Technology, and of Education, and organized by  {\em Instituto Nacional de Matem\'atica Pura e Aplicada} (IMPA). OBMEP has similar aims as the OBM but with exclusive focus on public schools, wherein the Brazilian educational system faces serious challenges. The organizers of the OBMEP promote different activities: the OBMEP Program in Schools, which is focused on the mathematics teachers by stimulating activities outside class hours; the Olympic Pole Intensive Training (POTI), which offers free math courses for students enrolled in the $8^{th}$ and $9^{th}$ grades of elementary school and in any year of high school interested in participating in the OBMEP and OBM; the {\em Scientific Initiation} Program (PIC-OBMEP) which aims at giving continuous support to OBMEP medalists through scholarships when they start studying in an university. These are some examples of  initiatives which are related to the organization of the OBMEP and are expected to strengthen the teaching of mathematics in public schools, awaken in students of public schools an interest for mathematics, and science in general, and provide those students who are OBMEP medalists with the opportunity to attend an university and build a career.

The OBMEP is held every year since  2005, when there were over 31,000 schools registered, comprising over 10.5 million students. In 2013 there were over 47,000 schools registered, involving nearly 19.2 million students, covering approximately 99,5\%  of the municipalities in Brazil. In 2013, OBMEP awarded 499 students with gold medals, 900 with silver, and 4,600 with bronze.

The OBMEP is structured as follows. The educational school  system in Brazil comprises 12 years of basic education, the first 9 years comprise the primary school and the remaining 3 are the secondary school.  Compared to other countries, the first 5 years can be compared to primary school, the next four grades can be compared to a low secondary school, and the last three grades are the secondary school or high school \citep{Biondi:Vasconcellos:Menezes:2012}.

The public Brazilian educational system comprises three different types of administrative school  levels: municipal, state, and federal. Any of these schools are allowed to subscribe to take part in the OBMEP. The registration is done by the schools, and each school indicates how many students will take part in the first phase of the OBMEP. The students are divided into three different levels:
\begin{itemize}
\item Level 1: students in the $6^{th}$ and $7^{th}$ grades of the primary school; 
\item Level 2: students in the $8^{th}$ and $9^{th}$ grades of the primary school;
\item Level 3: students in high school.
\end{itemize}
The OBMEP is performed in two phases: first, students take a multiple choice exam with 20 questions for each educational level. The correction of the first phase exams is done locally, that is, they are corrected by the school's own teachers. Approximately 5\% of students with the highest scores in each level of each school, are approved for the second phase of OBMEP. Students who scored zero are not qualified for the second phase, even if his/her school has not reached the proportion of  students expected to be in the second phase. In the second phase, students write a discursive examination comprising six questions. These tests are also separated by level of education. The exams of the second phase are marked regionally by committees formed by the OBMEP organizing committee. Typically, committee members are mathematical researchers from universities in the region, who have experience with Mathematics Olympiads. For every edition, the various regional committees define a cutoff point of the note. The marks are reviewed by a national committee who establishes the prizes to be awarded.

We aim at studying the performance of schools across Brazil that have taken part of the OBMEP from 2006  until 2013, the latest year that we have information available. Understanding what covariates influence the performance of schools in the OBMEP is important as it might help defining, or revising, strategies about teaching mathematics and attract more students to the area. 

This paper is organized as follows: next section describes the dataset available and how the sample to be analyzed was obtained. As we are comparing the performance of schools across different years, with different students being exposed to different exams, we propose a  standardization of the school's average scores, such that they lie in the interval $(0,1)$. For this reason, Section \ref{sec:review} shows a brief literature review on beta regression analysis. And Section \ref{sec:model} proposes a hierarchical dynamic  beta regression model to analyze the performance of schools across the OBMEP editions from 2006 until 2013. The inference procedure is also described in detail in this section. Then, Section \ref{sec:result}  describes the model comparison criteria used to choose the best model among those fitted and discusses the results under this best model. Finally, Section \ref{sec:conclusion} discusses our findings and describes some possible avenues of future research about the OBMEP.
 
\section{Dataset description \label{sec:dataprocess}} 

We have information available from three different sources. The organizers of the OBMEP provided us with all information from all students who registered for the OBMEP from 2005 until 2013.  We decided for removing the year of 2005 from the analysis because there are no records on the gender of the students. This information started to be collected from 2006 on. Table \ref{tab:freqschools} shows the number of schools that registered for each phase of the OBMEP from 2006 until 2013. Although the number of students present in the second phase reduces considerably when compared to that of the first phase, the proportion of schools  in both phases tend to be around 90\% every year.
\begin{table}[!hbt]
\centering
\begin{tabular}{|c|ccc|} \hline
Year & \multirow{1}{*}{No. of Schools} & \multirow{1}{*}{No. of Schools}  & \multirow{1}{*}{Percentage of schools}  \\ 
 & in Phase 1 & in Phases 1 and 2 &  in both phases  \\ \hline
2006 & 32,603 & 29,660 & 91.0\% \\
2007 & 37,886 &  35,480 &  93.6\% \\
2008 & 40,396 &  35,913 &  88.9\% \\ 
2009 & 43,851 &  39,379 &  89.8\% \\
2010 & 44,718 & 39,931 & 89.3\% \\
2011 & 44,684 & 39,928 & 89.4\% \\
2012 & 46,722 &  40,804 &  87.3\% \\
2013 & 47,145  & 42,483 &  90.1\% \\ \hline
\end{tabular}
\caption{Distribution of the number of schools registered for each edition and phase of the OBMEP from 2006 until 2013.\label{tab:freqschools}}
\end{table}
For each year, we have information on the performance of each student within each school in both phases. We also have the name and the national code of the schools. These can be linked to the schools' census data, which is collected nationwide, every year, by {\em Instituto Nacional de Estudos e Pesquisas Educacionais An\'isio Teixeira} (INEP, {\tt http://portal.inep.gov.br/}). The census data have information about local characteristics of the schools. Previous studies about the OBMEP have suggested that the performance of students is strongly related to the geographical region the schools are located in. As in Brazil the geographical regions are strongly related to the human development index (HDI), we also obtained information about the HDI in 2010 of each Brazilian municipality present in the data. This is available from {\tt http://www.pnud.org.br/IDH/DH.aspx}.

This initial study focuses on the average scores of the schools that took part in the second phase of the OBMEP.  To ease the computational burden of estimating our models we choose to analyze a sample from this population. Next we discuss how this sample was obtained. 

\subsection{Sampling design}

The locations of the schools that take part in the OBMEP are divided between urban and rural areas. In 2013,  70.1\% of the schools that participated in the OBMEP are located in urban areas, among these, 0.6\% are federal, 42.8\% are state and 26.6\% are municipal. The remaining 0.1\%  are private  schools that incorporate some students from the public system and offer  a curriculum similar to  the public one. These private schools are excluded from this study. Throughout the years the distribution of urban and rural schools taking part in the OBMEP follows similar patterns. As the rural schools involve too many particularities we opted to focus only on  schools located in urban areas. 

Our aim is to model the average score of the schools in the second phase of the OBMEP. In order to minimize the variance of the mean average scores we propose a stratified random sampling scheme \citep{Thompson:2012}. The strata are defined by the following three auxiliary variables: the educational level (1, 2, and 3), the administrative level of the school (federal, state or municipal), and different levels of the HDI.  The behavior of the HDI across Brazil is strongly related to the geographical regions\footnote{See e.g. https://en.wikipedia.org/wiki/List\_of\_Brazilian\_federative\_units\_by\_Human\_Development\_Index .}, assuming high values in the south, and smaller values in the north and north-east regions of the country. We expect this will capture local characteristics of where the school is located in. We assume $z=HDI \in (0,1)$ with probability density function $f(z)$. Let $z_0$ and $z_U$ be the smallest and largest values of $z$ in the population. We obtain stratum boundaries, $z_1,z_2,\cdots, z_{U-1}$, by minimizing $V(\overline{z})=\frac{1}{n}\sum_{h=1}^U W_h S_h^2$ and ignoring the finite population correction factor \citep{Dalenius:Hodges:1959}. In the previous equation, $W_h=N_h/N$ is the stratum weight, and $S_h^2$ is the true variance of the stratum. Following this procedure, HDI was divided into 5 categories. When the ranges of the three auxiliary variables are combined 45 strata result. However, as federal schools tend to perform best when compared to other type of schools, and they represent only $0.6\%$ of the schools in urban areas, we decided to define federal schools as a certainty strata, leading to 31 strata in total.

The sample was obtained by first selecting nearly 20\% of the schools that took part in the  2006 edition of the OBMEP. From 2006 onwards, we checked the schools that took part in the subsequent editions and only kept in the sample those which participated in all editions until 2013. The final sample size comprises $n=2,463$ schools.

\paragraph{Standardization of the schools average score \label{sec:standard}}

We propose a standardization of the average scores of the schools such that they are comparable across years. Let $W_{ijt}$ be the average score of school $j$ within level $i$ in year $t$, $i=1,2,3$, $j=1,2,\cdots,n_i$, $t=1,2,\cdots,8$. Define 
$\overline {W}_{it} = \frac{\sum_{j}W_{ijt}}{n_{i}}$ as the average score of all schools within level $i$ in year $t$. Now define  $Z_{ijt} = \frac{W_{ijt}-\overline {W}_{it}}{S_{it}}$
where $S_{it}$ is the standard deviation of $W_{ijt}$ in year $t$ and level $i$. As the average scores $W_{ijt}$ fall in the interval $(0,120)$, we then compute $min_{it} = \frac{0 - \overline {W}_{it}}{S_{it}}$ and $max_{it} = \frac{120- \overline {W}_{it}}{S_{it}}$ to obtain $Y_{ijt} = \frac{ Z_{ijt}- min_{it}}{max_{it}-min_{it}}$, such that $Y_{ijt}  \in (0,1)$. This is the quantity that will be considered as response variable in the models that will be fitted in Section \ref{sec:result}.


\subsection{A brief literature review \label{sec:review}}

The use of the beta distribution to model rates and proportions as a function of covariates is relatively recent in the literature. \cite{Ferrari:Cribari:2004} proposed a beta regression model for rates and proportions. They provide closed-form expressions for the score function, for the Fisher's information matrix and perform hypothesis testing of the coefficients using approximations based on the asymptotic normality of the maximum likelihood estimator. 
In particular,  \cite{Ferrari:Cribari:2004}, assume that if $Y  \sim beta(\mu,\phi)$ then $f(y\mid \mu,\phi)=\frac{\Gamma(\phi)}{\Gamma(\mu\phi)\Gamma((1-\mu)\phi)} y^{\mu\phi-1}(1-y)^{((1-\mu)\phi-1)}$, such that $E(Y)=\mu$ and $Var(Y)=\frac{\mu(1-\mu)}{1+\phi}$, for $y \in (0,1)$, $\mu \in (0,1)$, and $\phi>0$. They focus on the modelling of a transformation of $\mu$ as a function of covariates.

\cite{Branscum:Johnson:Thurmond:2007} discuss beta regression from a Bayesian point of view. In their model, the mean depends on covariates through a logistic link function. They also propose  a semiparametric beta regression, and model fitting is performed using {\tt WinBUGS} \citep{Lunn2000}.

\cite{Silva:Migon:Correia:2011} develop a Bayesian dynamic beta regression model for time series of rates or proportions. They propose to approximate the posterior distribution of the state parameters through Bayesian linear estimation and Gaussian quadrature, avoiding the use of Markov chain Monte Carlos (MCMC) methods.

\cite{Bayes:Bazan:Garcia:2012} propose a beta rectangular regression model which allows more flexibility in the modelling of  the tails  and of the precision parameter when compared to the beta regression model. The inference procedure follows the Bayesian paradigm and they also use the software {\tt WinBUGS} to obtain samples from the resultant posterior distribution of the model parameters.

As the data described in Section \ref{sec:dataprocess} has a natural hierarchical structure, in the next section we propose a hierarchical dynamic beta regression model that naturally accounts for the different educational levels as well as the evolution in time of the observations. Also, we allow the precision parameter of the beta distribution to be a function of covariates, possibly different from the ones in the mean structure.

\section{Proposed model \label{sec:model}}

Let $Y_{ijt}$ be the average score of school $j=1,2,\cdots,n_i$,  within educational level $i=1,2,\cdots,I$, in year $t=1,2,\cdots,T$.
As described in Section \ref{sec:standard} the average scores of the schools were standardized within each year such that $Y_{ijt}$ is a random variable defined in the interval $(0,1)$. In particular, we assume
\begin{eqnarray*}
(Y_{ijt}\mid \mu_{ijt},\phi_{ijt}) \sim beta(\mu_{ijt},\phi_{ijt}),
\end{eqnarray*}
follows a beta distribution with mean $\mu_{ijt}$, $0<\mu_{ijt}<1$ that represents the average score of school $j$ within level $i$ in year $t$, and  $\phi_{ijt}>0$ can be seen as a precision parameter \citep{Ferrari:Cribari:2004}. In what follows we describe the proposed modelling of the components $\mu_{ijt}$ and $\phi_{ijt}$.

Let ${\bf X}_{ijt}$ be a $p$-dimensional vector of covariates, and ${\bfbeta}_{it}$ a $p$-dimensional vector of coefficients, ${\bfbeta}_{it}=(\beta_{0it}, \cdots, \beta_{(p-1)it})'$, we assume that
\begin{subequations} \label{eq:meanall}
\begin{align} 
\log \left(\frac{\mu_{ijt}}{1-\mu_{ijt}}\right)&={\bf X}_{ijt}' {\bfbeta}_{it}, \mbox{ with } \label{eq:mean} \\
{\bfbeta}_{it}&={\bfalpha}_{t}+{\bf v}_{it}, \quad {\bf v}_{it} \sim N(0,V_{\beta i}), \, t=1,\cdots, T,  \label{eq:beta}\\
{\bfalpha}_{t}&={\bfalpha}_{t-1}+{\bfomega}_t, \quad {\bfomega}_t \sim N(0,W_{\alpha}). \label{eq:evolution}
\end{align}
\end{subequations}
The components ${\bf v}_{it}$ and ${\bfomega}_t$ are assumed mutually and internally independent, for all $i$ and $t$.
Note that the coefficients of the covariates in equation \eqref{eq:mean} vary with the educational level $i$, and year $t$. Also, {\em a priori}, the coefficients ${\bfbeta}_{it}$  follow a hierarchical dynamic model \citep{gamerman:migon:1993}, in the sense that for each time $t$, and level $i$, ${\bfbeta}_{it}$ has mean ${\bfalpha}_t$ which in turn evolves smoothly with time according to equation \eqref{eq:evolution}.
The parameter vector ${\bfalpha}_t=(\alpha_{0t},\cdots,\alpha_{(p-1)t})'$ is a $p-$dimensional  vector, with each component representing the overall effect of the $l^{th}$ covariate on the logit of the mean  $\mu_{ijt}$. The variances of the prior distribution of the coefficients, $V_{\beta i}$,  also vary with the educational level $i$, such that $V_{\beta i}$ is a $p$-dimensional diagonal matrix, with elements $V_{\beta i m}$, $m=0,1,2,\cdots,p-1$. And $W_{\alpha}$ is a $p$-dimensional diagonal matrix, with each element of the diagonal representing the variance of the evolution in time of component $\alpha_{lt}$, $l=1,\cdots,p$.

\begin{figure}[!hbt]
\begin{center}
\includegraphics[scale=0.8,angle=0]{\dir 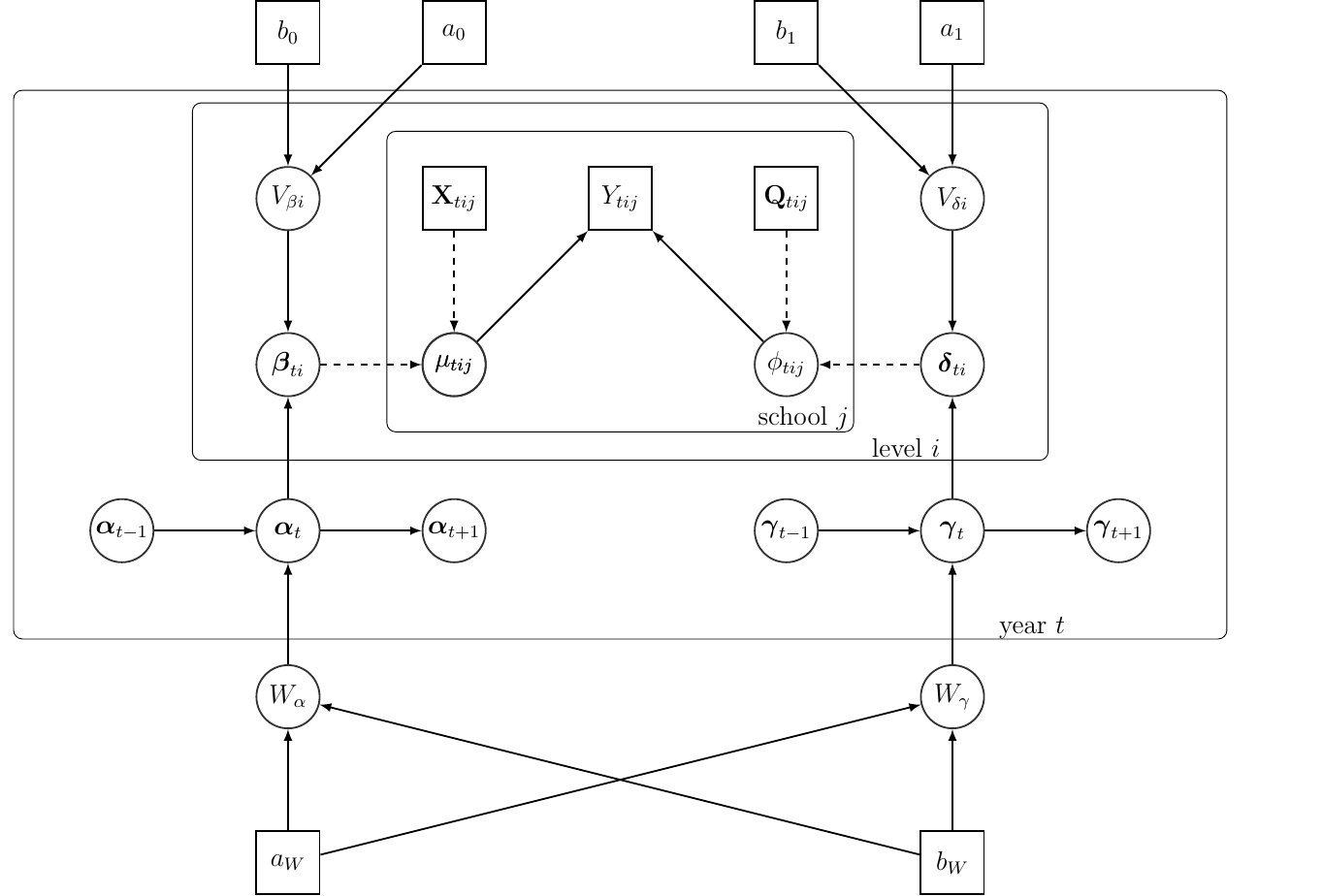}
\caption{Directed acyclic graph of the hierarchical model proposed in equations (\ref{eq:meanall}) and (\ref{eq:precall}).\label{fig:dag} }
\end{center}
\end{figure}

For the precision parameter $\phi_{ijt}$ we assume
\begin{subequations} \label{eq:precall}
\begin{align} 
\log \phi_{ijt}&=-{\bf Q}_{ijt}' {\bfdelta}_{it}, \mbox{ with } \label{eq:precision} \\
{\bfdelta}_{it}&={\bfgamma}_{t}+{\bf v}_{1it}, \quad {\bf v}_{1it} \sim N(0,V_{\delta i}), \, t=1,\cdots, T, \label{eq:delta}\\
{\bfgamma}_{t}&={\bfgamma}_{t-1}+{\bfomega}_{1t}, \quad {\bfomega}_{1t} \sim N(0,W_{\gamma}), \label{eq:evolutionprec}
\end{align}
\end{subequations}
where ${\bf Q}_{ijt}$ is a $q$-dimensional vector containing the covariates that might affect the precision parameter $\phi_{ijt}$, and  ${\bfdelta}_{it}$ is a $q-$dimensional vector of coefficients that,  {\em a priori}, also follow a hierarchical dynamic structure. Note that ${\bfgamma}_t$ is a $q$-dimensional vector, such that each component captures the overall mean of the respective component in ${\bfdelta}_{it}$. As $\phi_{ijt}$ is a precision parameter, we use a negative sign in equation \eqref{eq:precision} to ease interpretation of the coefficients ${\bfdelta}_{it}$ \citep{Smithson:2006}. The $q-$dimensional covariance matrix $V_{\delta i}$ is allowed to vary per level and is assumed  to be diagonal, $W_{\gamma}$ is also a diagonal matrix, implying prior independence among the components of ${\bfdelta}_{it}$ and ${\bfgamma}_t$,  respectively. Figure \ref{fig:dag} depicts a directed acyclic graph of the proposed model.

\subsection{Prior specification and inference procedure}

Inference procedure is performed under the Bayesian paradigm. To complete model specification and following equations \eqref{eq:meanall} and \eqref{eq:precall}, we are left to assign the prior distribution of the hypeparameters $V_{\beta i}$, $V_{\delta i}$, for $i=1,2,\cdots,I$, $W_{\alpha}$, $W_{\gamma}$, ${\bfalpha}_0$, and ${\bfgamma}_0$. We assume prior independence among the hyperparameters. For the  variance parameters, we assign independent, inverse gamma prior distributions with infinite variance and prior mean fixed at some reasonable value,  e.g. the maximum likelihood estimate based on independent fits for each year.  For the components of ${\bfalpha}_0$ and ${\bfgamma}_0$ we assign independent, zero mean normal distributions, with variance fixed at some reasonably large value.

Let ${\bf y}$ be the vector comprising the average scores of the schools stacked across the different educational levels and  years. And let ${\bfTheta}$ be the parameter vector comprising all the parameters and hyperparameters in equations \eqref{eq:meanall} and \eqref{eq:precall}. The likelihood function, $f({\bf y} \mid {\bfTheta})$, is given by
$$
f({\bf y} \mid {\bfTheta})=\prod_{t=1}^T \prod_{i=1}^I \prod_{j=1}^{n_{i}}  \frac{\Gamma(\phi_{ijt})}{\Gamma(\mu_{ijt}\phi_{ijt})\Gamma\left((1-\mu_{ijt})\phi_{ijt}\right)} y_{ijt}^{[\mu_{ijt}\phi_{ijt}-1]}\left(1-y_{ijt}\right)^{[(1-\mu_{ijt})\phi_{ijt}-1]},
$$
where $\Gamma(.)$ is the usual Gamma function.

Following the Bayes' theorem, the posterior distribution of ${\bfTheta}$, $p({\bfTheta} \mid {\bf y})$, is proportional to the likelihood function times the prior distribution. As we assume independence among the hyperparameters, it follows that
{\footnotesize
\begin{eqnarray*}
\hspace{-0.8cm} &&p({\bfTheta} \mid {\bf y}) \propto f({\bf y} \mid {\bfTheta}) \left\{ \prod_{i=1}^I \left\{\prod_{t=1}^T \left[ p({\bfbeta}_{it} \mid {\bfalpha}_t, V_{\beta i}) p({\bfalpha}_t \mid {\bfalpha}_{t-1},W_{\alpha}) \right] \left[ p({\bfdelta}_{it} \mid {\bfgamma}_t, V_{\delta i}) p({\bfgamma}_t \mid {\bfgamma}_{t-1},W_{\gamma}) \right] \right\} \right.\\
&&\qquad \qquad \qquad \left.\left[\prod_{m=0}^{p-1} p(V_{\beta im }) \, p(W_{\alpha m}) \right] \left[\prod_{k=0}^{q-1} p(V_{{\delta} ik}) \, p(W_{\gamma k})\right] \right\}   p({\bfalpha}_0 \mid {\bf m}_0,C_0) p({\bfgamma}_0 \mid {\bf m}_0,C_0)    ,
\end{eqnarray*}
}
which does not have a closed analytical form. We make use of MCMC  methods to obtain samples from the posterior distribution above. In particular we use a hybrid Gibbs sampler  with some steps of the Metropolis-Hastings algorithm. The posterior full conditional distributions of  ${\bfbeta}_{it}$ and ${\bfdelta}_{it}$ do not have a closed form, and are sampled using the Metropolis-Hastings algorithm. In particular, the MCMC algorithm is implemented using the {\tt JAGS} software \citep{Plummer:2003}.

\section{Data Analysis  \label{sec:result}}

In this Section we analyze the performance of the schools across the different editions of the OBMEP, from 2006 until 2013. 
Equations \eqref{eq:meanall} and \eqref{eq:precall} propose the most general model specification for analyzing the data described in Section \ref{sec:dataprocess}. We fit particular cases of the proposed model and use three model comparison criteria to choose the best model among those fitted. All fitted models assume the  mean structure $\mu_{ijt}$ as a function of an intercept, and the following covariates: the school's administrative level (ADM), with ADM=1 if the school is federal, and $0$ otherwise, the standardized human development index of the municipality the school is located in (HDI), the presence of library (LIB), the presence of laboratory (LAB), and the standardized proportion of boys present in the second phase of the OBMEP (BOYS). Note that the HDI is a proxy to describe the social condition that schools located in the same municipality share. Therefore, ${\bf X}_{ijt}=(1,ADM,HDI, LIB,LAB,BOYS)$, and ${\bfbeta}_{it}=(\beta_{0it},\beta_{1it},\beta_{2it},\beta_{3it},\beta_{4it},\beta_{5it})'$.

For the precision parameter we explore different models, varying from a constant precision parameter for each level, to different versions that assume the logarithm of the precision $\phi_{ijt}$ as a linear function of the number of students (denoted as $nstudent$) who were present in the second phase of the OBMEP. This allows the precision parameter of the beta distribution to change with the number of students the school has, in each level, in the second phase of the OBMEP.  The fitted models consider particular versions of equations (\ref{eq:precision}) and (\ref{eq:delta}); in particular,  the following models are fitted:
\begin{itemize}
\item[M1] $Q_{ijt}=1$, $\delta_{it}=\delta_{0}$, $\forall t=1,2,\cdots,T$ and $i=1,2,3$,  $q=1$;
\item[M2] ${\bf Q}_{ijt}'=(1,nstudent_{ijt})$ ${\bfdelta}_{it}=(\delta_{0},\delta_{1})'$, $\forall t=1,2,\cdots,T$, and $i=1,2,3$, $q=2$;
\item[M3] ${\bf Q}_{ijt}'=(1,nstudent_{ijt})$ ${\bfdelta}_{it}=(\delta_{0i},\delta_{1i})'$, $\forall t=1,2,\cdots,T$, $q=2$;
\item[M4] ${\bf Q}_{ijt}'=(1,nstudent_{ijt})$, ${\bfdelta}_{it}=(\delta_{0t},\delta_{1t})'$, for $i=1,2,3$, $q=2$;
\item[M5] ${\bf Q}_{ijt}'=(1,nstudent_{ijt})$, ${\bfdelta}_{it}=(\delta_{0it},\delta_{1it})'$, $q=2$.
\end{itemize}
Note that M1 assumes the precision fixed across different levels and years, M2 describes the logarithm of the precision as a linear function of an intercept and the number of students, per school, present in the second phase of the OBMEP. Models  M3 and M4 allow the coefficients to vary by level or year, and M5 allows the coefficients to vary by level and year simultaneously,  corresponding to the most general  proposed model in equation (\ref{eq:precall}).

 For each model we let the MCMC run for 35,000 iterations, used 5,000 as burn in and stored every 30th iteration.   Convergence was checked using the diagnostic tools in the {\tt R} package {\tt coda} \citep{plummer:2006}.

\subsection{Model comparison}

In this Section we describe the different model comparison criteria used to compare the different fitted models. In particular we use the deviance information criterion proposed by \citep{spiegelhalter:best:carlin:linde:2002}, and two other criteria  based on proper scoring rules.

\paragraph{Deviance Information Criterion ($DIC$)} The $DIC$ is a generalization of the AIC based on the posterior distribution of the deviance, $D({{\bfTheta}})=-2\log \, p( {\bf y} \mid {\bfTheta})$ \citep{spiegelhalter:best:carlin:linde:2002}. More formally, the DIC is defined as 
$$
DIC=\overline{D}+p_D=2\overline{D}-D(\overline{\bfTheta}),
$$
where $\overline{D}$ defines the posterior expectation of the deviance, $\overline{D}=E_{{\bfTheta} \mid {\bf y}}(D)$, $p_D$ is the effective number of parameters, with $p_D=\overline{D}-D(\overline{\bfTheta})$, and $\overline{\bfTheta}$ represents the posterior mean of the parameters. $\overline{D}$ might be seen as a goodness of fit measurement, whereas $p_D$ indicates the complexity of the model. Smaller values of $DIC$ indicate better fitting models. 

\paragraph{Scoring rules}
\cite{gneiting:raftery:2007} consider proper scoring rules for assessing the quality of probabilistic forecasts.  Following \cite{gsch:czad:2008}, we use the same data for estimation and computation of the scores, as our focus is on understanding the relationship between the schools' performance and the covariates other than prediction.  We use two different scoring rules: 

\paragraph{Ranked probability score ($RPS$)} For each $y_{ijt}$, the RPS can be expressed as
$$
RPS(y_{ijt}) = E|y^{rep}_{ijt} - y_{ijt}| - \frac{1}{2} E|y^{rep}_{ijt} - \tilde{y}^{rep}_{ijt}|,
$$
where $y_{ijt}$ is the observed average score of the $j^{th}$ school within level $i$ in year $t$, $y^{rep}_{ijt}$ and $\tilde{y}^{rep}_{ijt}$ are independent replicates from the posterior predictive distribution of the respective model. 

Assuming there is a sample of size $L$ from the posterior distribution of the parameters in the model, we can  obtain roughly independent replicates, $y^{rep}_{ijt}$ and $\tilde{y}^{rep}_{ijt}$, from the respective posterior predictive distribution. The components $E|y^{rep}_{ijt} - y_{ijt}|$ and $E|y^{rep}_{ijt} - \tilde{y}^{rep}_{ijt}|$ can be approximated using Monte Carlo integration through
$\frac{1}{L} \sum_{l=1}^L |y^{{rep}^{(l)}}_{ijt} - y_{ijt}|$ and 
$\frac{1}{L} \sum_{l=1}^L |y^{{rep}^{(l)}}_{ijt} - \tilde{y}^{{rep}^{(l)}}_{ijt}|$, and 
$$
RPS=\frac{1}{n}\sum_{i=1}^I \sum_{j=1}^{n_{i}} \sum_{t=1}^T RPS(y_{ijt}),
$$
where $n$ is the total number of schools across all the years and educational levels.
Smaller values of  $RPS$ indicate the best model among the fitted ones.

\paragraph{Logarithmic score ($LogS$)}
The logarithmic score is defined as $- \log p(y_{ijt})$, where $p(y_{ijt})$ is the probability density function at the observed average score of  school $j$ in the $i^{th}$ level and year $t$. Considering the observed sample ${\bf y}$, $LogS$ is computed as 
$$
LogS=\frac{1}{n}\sum_{i=1}^I \sum_{j=1}^{n_{i}} \sum_{t=1}^T -\log \, p(y_{ijt}),
$$
where $n$ is the total number of schools across all the years and educational levels.
Smaller values of  $LogS$ indicate the best model among the fitted ones.

Assuming there is a sample from the posterior distribution of the parameters of size $L$ available, the predictive distribution $p(y_{ijt})$ is approximated using Monte Carlo integration, that is, 
$$
p(y_{ijt})=\int_{\bfTheta}p_y(y_{ijt} \mid {\bfTheta}) \pi(\Theta \mid y) d{\bfTheta} \, \approx \, \frac{1}{L}\sum_{l=1}^L p_y(y_{ijt} \mid {\bfTheta}^{(l)}),
$$
where $p_y(y_{ijt} \mid {\bfTheta}^{(l)})$ is the probability density function of the beta distribution conditioned on the $l^{th}$ sampled value of the parameter vector ${\bfTheta}$, evaluated at $y_{ijt}$.  

Table \ref{tab:comparison} shows the values of the different model comparison criteria obtained under each fitted model. Although the values of $LogS$ are quite similar across the different models, M5 performs slightly better. Model M5 performs best under $DIC$ and $RPS$. The results we show next are based on those obtained  under model M5.

\begin{table}[h]
\centering
\begin{tabular}{lccccc } \hline
Model & $\overline{D}$ & $p_D$ & $DIC$ & $RPS$ & $LogS$\\  \hline
M1  & -56247.15 & 167.36 & -56079.79   & 0.03653 & -1.43\\
M2  & -57126.82  & 155.48 & -56971.33  & 0.03637  & -1.45 \\
M3  & -57322.07  & 177.92 & -57156.78  & 0.03635  & -1.45 \\
M4 & -57471.56  & 170.94 & -57300.61   & 0.03620  & -1.46 \\
M5 & -58031.94 & 209.54 & {\em -57822.40}    & {\em 0.03610}  & {\em -1.47}  \\ \hline
\end{tabular}
\caption{Model comparison criteria, $DIC$ and its components ($p_D$ and $\overline{D}$), $RPS$, and $LogS$, under each fitted model. Numbers in italics indicate best model under the respective criterion. \label{tab:comparison}}
\end{table}

\paragraph{Fitting a normal model to $logit \, y_{ijt}$}

\begin{figure}[!hbt]
\begin{center}
\includegraphics[scale=0.4]{\dir 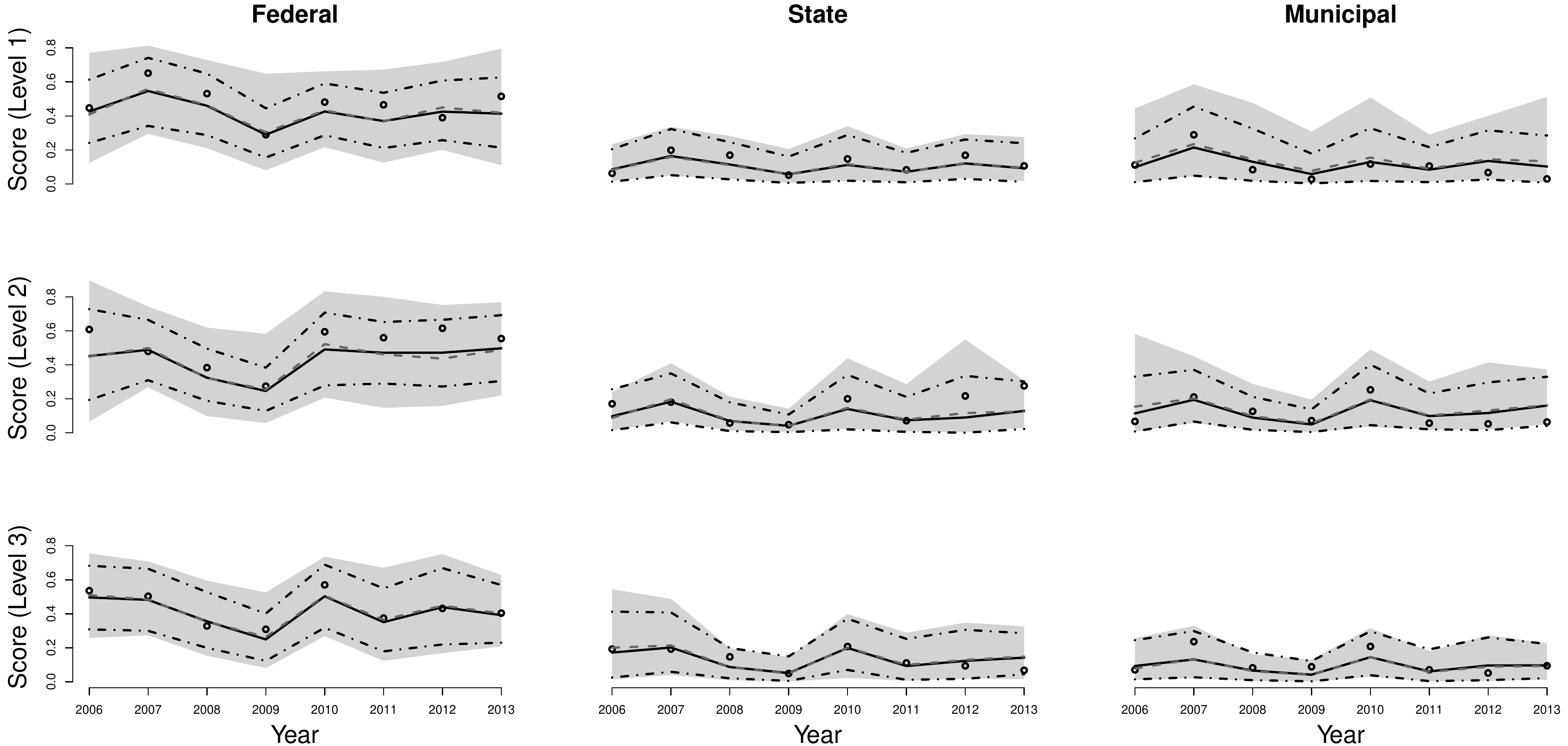}
\caption{Posterior summary of the predictive distribution of three different schools according to their administrative (columns) and educational levels (rows), together with the observed values (open circles). The gray shaded area is the 95\% posterior predictive credible interval for  $\hat{y}_{ijt}=\frac{\exp(y^*_{ijt})}{1+\exp(y^*_{ijt})}$, the gray dashed line is the mean of the posterior predictive distribution of $\hat{y}_{ijt}$. The black solid line is the posterior mean of the predictive distribution, and the dotted-dashed lines are the limits of the 95\% posterior predictive credible interval under the beta hierarchical model M5. \label{fig:prediction} }
\end{center}
\end{figure}

The standardization of the observations to the interval $(0,1)$ turns it possible a comparison of the scores of the  schools across the different years and levels. An alternative to the beta hierarchical regression model is to fit a normal hierarchical model to $y^*_{ijt}=logit \, y_{ijt}$. Basically, this assumes that $y^*_{ijt} \sim N(\mu_{ijt},\sigma^2_{ijt})$, with $\mu_{ijt}={\bf X}_{ijt}' {\bfbeta}_{it}$, and $\log \phi_{ijt}={\bf Q}_{ijt}' {\bfdelta}_{it}$, where $\phi_{ijt}=\frac{1}{\sigma^2_{ijt}}$. The evolution in time of parameters ${\bfbeta}_{it}$  and ${\bfdelta}_{it}$ follow the same dynamic structure as in equations (\ref{eq:beta}) and (\ref{eq:delta}), respectively. This model was fitted to the data following similar prior specifications as those used when fitting  model M5 under the beta response structure. The values of the different model comparison criteria were the following: DIC = 48026.54, RPS = 0.4547, and logS = 1.21, all of them much greater than the respective values obtained under model M5, suggesting that it is better to fit the beta hierarchical regression model  which considers the original scale of the observations. Another interesting comparison is to look at the fitted values under model M5 and this normal counterpart. To make this comparison we first obtained samples from the posterior predictive distribution under the beta hierarchical model M5 and from the normal hierarchical model. Then, under the normal model, we transformed the fitted values back to the original scale to be able to compare the results from the normal model with those from M5.
From panels of Figure \ref{fig:prediction} it is clear that both models provide similar values of the means of the posterior predictive distribution. However, the beta hierarchical model provides ranges of the 95\% posterior predictive interval which are, in general, narrower than the ones obtained under the normal model. Similar results were obtained for other schools in the sample. 

Next we focus on the description of the posterior distribution of the parameters in model M5 to better understand the effect of covariates on the performance of schools across years and levels.

\begin{figure}[!hbt]
\begin{center}
\subfigure{\includegraphics[scale=0.3]{\dir 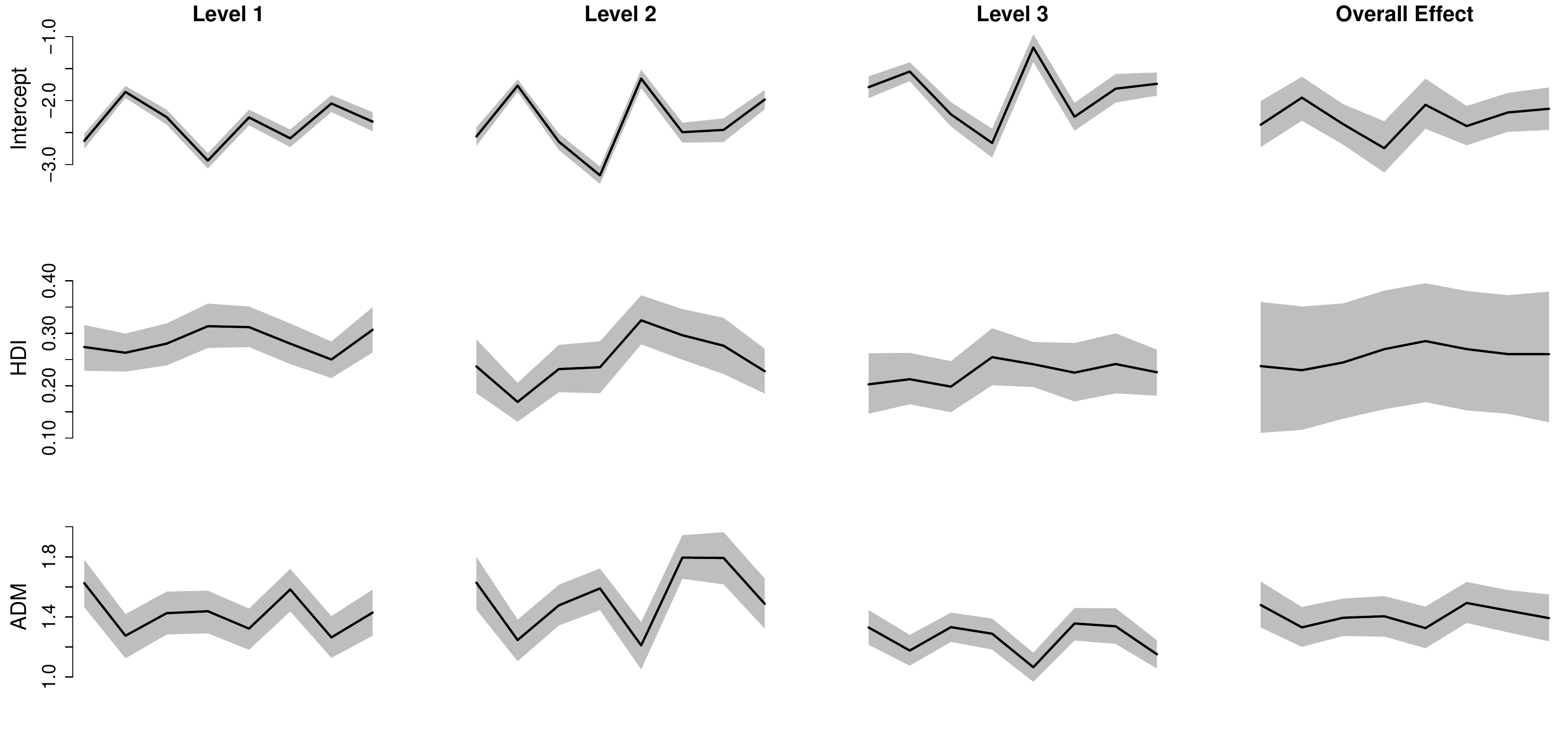}} \\
\subfigure{\includegraphics[scale=0.3]{\dir 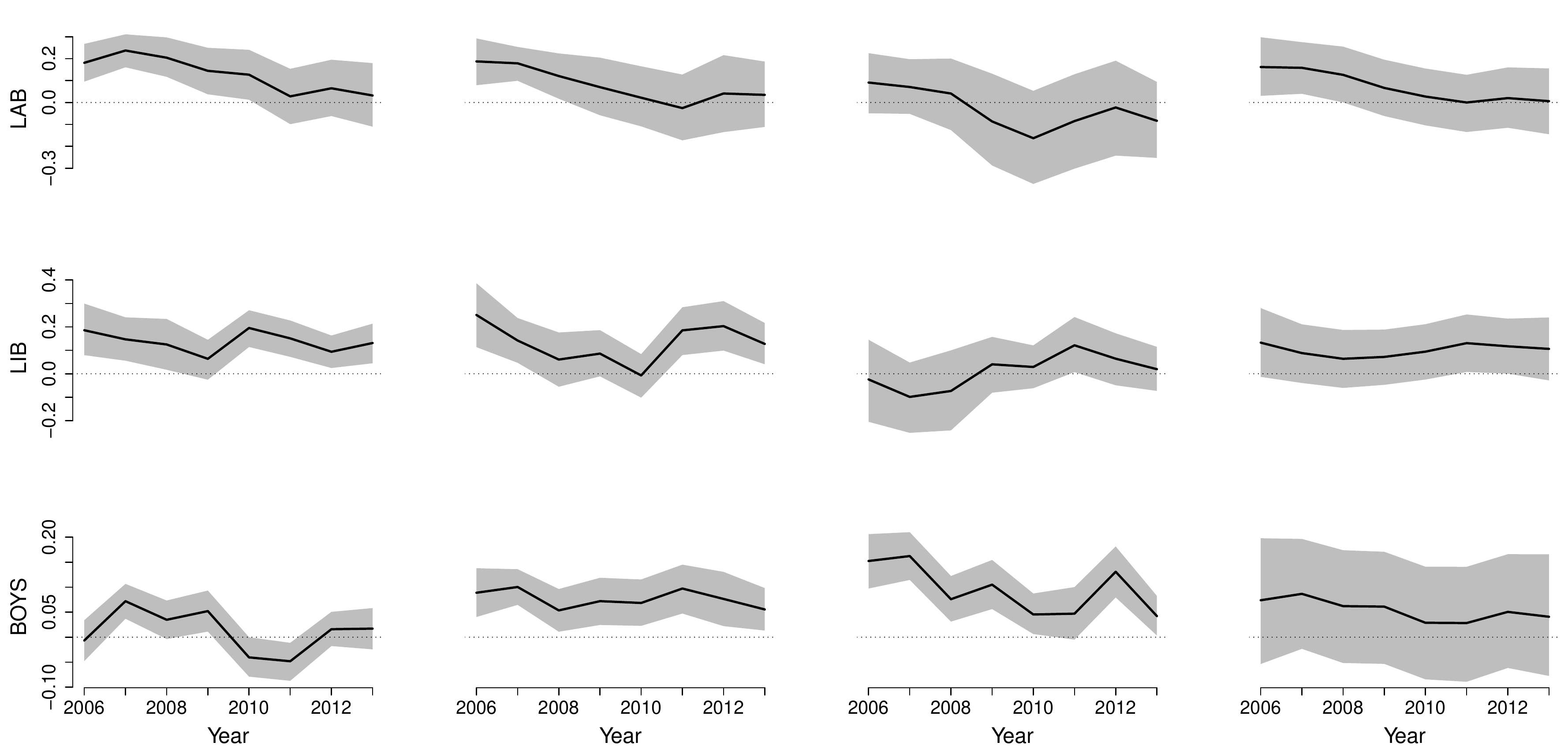}}
\caption{Posterior summary (mean: solid line, and limits of the 95\% credible intervals: shaded area) of the coefficients $\beta_{lit}$, for the intercept, HDI, ADM, LAB, LIB, and BOYS (rows) by level (columns 1, 2 and 3), and respective overall effect, $\alpha_{lt}$ ($4^{th}$ column), $l=0,1,2,3,4,5$.\label{fig:coefficients} }
\end{center}
\end{figure}

The intercept per level, $\beta_{0it}$, suggests some cyclical pattern in the performance of the schools across the levels and years. For level 1 this pattern is smoother than for levels 2 and 3. In particular, in 2009, levels 2 and 3 show a drop in the performance, followed by an increase in 2010, suggesting that the exam in 2009 resulted in the lowest scores among the editions of the OBMEP considered in the sample. After 2010, the scores show a slight increase for levels 2 and 3 (first row of Figure \ref{fig:coefficients}). Indeed, most of the schools commented with the organizers of the OBMEP that the exam in 2009 was too difficult when compared to previous years.

As expected, regardless of the level considered, an increase in the value of the HDI impacts positively the logit of the mean performance of a school in the second phase of the OBMEP. The estimated overall effect of the HDI does not show any particular pattern across the years, suggesting a constant positive effect across years (second row of Figure \ref{fig:coefficients}). This is expected as we are using the value of the HDI in 2010 for all years.

Regarding the administrative level, federal schools perform considerably better than state or municipal ones. The evolution of the effect of the administrative level across the years seems to be relatively similar for levels 2 and 3, with both being slightly different from level 1, especially after 2010. The overall effect of the administrative level shows a slight increase after 2010 (third row of Figure \ref{fig:coefficients}). Also, the effect of the administrative level of the school tends to be smaller for level 3 than for levels 1 and 2, especially after 2010.

The presence of a laboratory does not seem to affect the logit of the overall mean for level 3, whereas for levels 1 and 2 it has a small, positive effect until 2009 and 2007, respectively, with $0$ falling within the 95\% credible interval after these years. A similar behaviour is observed for the coefficient of the presence of a library in the school ($4^{th}$ and $5^{th}$ rows of Figure \ref{fig:coefficients}).

Although the 95\% posterior credible interval of the overall effect of the proportion of boys in the second phase includes  $0$ for all years ($6^{th}$ row and $4^{th}$ column  of Figure \ref{fig:coefficients}), interesting characteristics of the effect of this covariate are observed when we disentangle the effect per level. For levels 2 and 3, an increase in the proportion of boys result in an increase in the respective logit of the mean score of the school. On the other hand, the proportion of boys in the second phase does not seem to affect  the logit of the mean score of level 1, being strictly positive only in 2007.

\begin{figure}[!hbt]
\begin{center}
\includegraphics[scale=0.30]{\dir 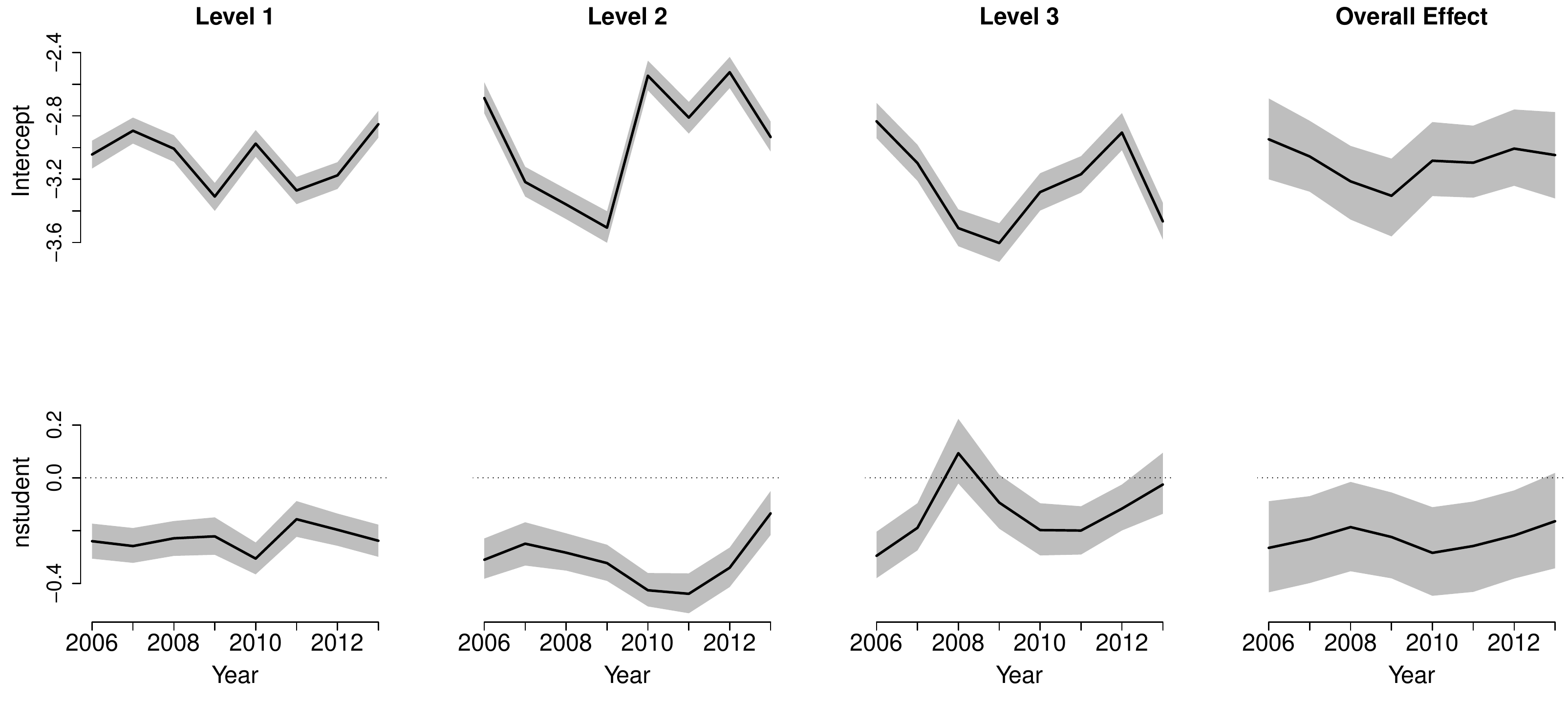}
\caption{Posterior summary (mean: solid line, and limits of the 95\% credible intervals: shaded area) of the intercept and the coefficient of $nstudent$ for the precision parameter, $\delta_{mit}$, together with the respective overall effect, $\gamma_{mt}$, $m=0,1$ ($4^{th}$  column).\label{fig:coefficientsprec} }
\end{center}
\end{figure}

The posterior summary of the coefficients related to the modelling of the precision parameter confirm the need of allowing different values across years and levels (see panels of Figure \ref{fig:coefficientsprec}). An increase in the number of students present in the second phase of OBMEP result in an increase in the respective precision parameter. Also, these effects vary across levels and years.

\begin{figure}[!htb]
\begin{center}
\includegraphics[scale=0.38]{\dir 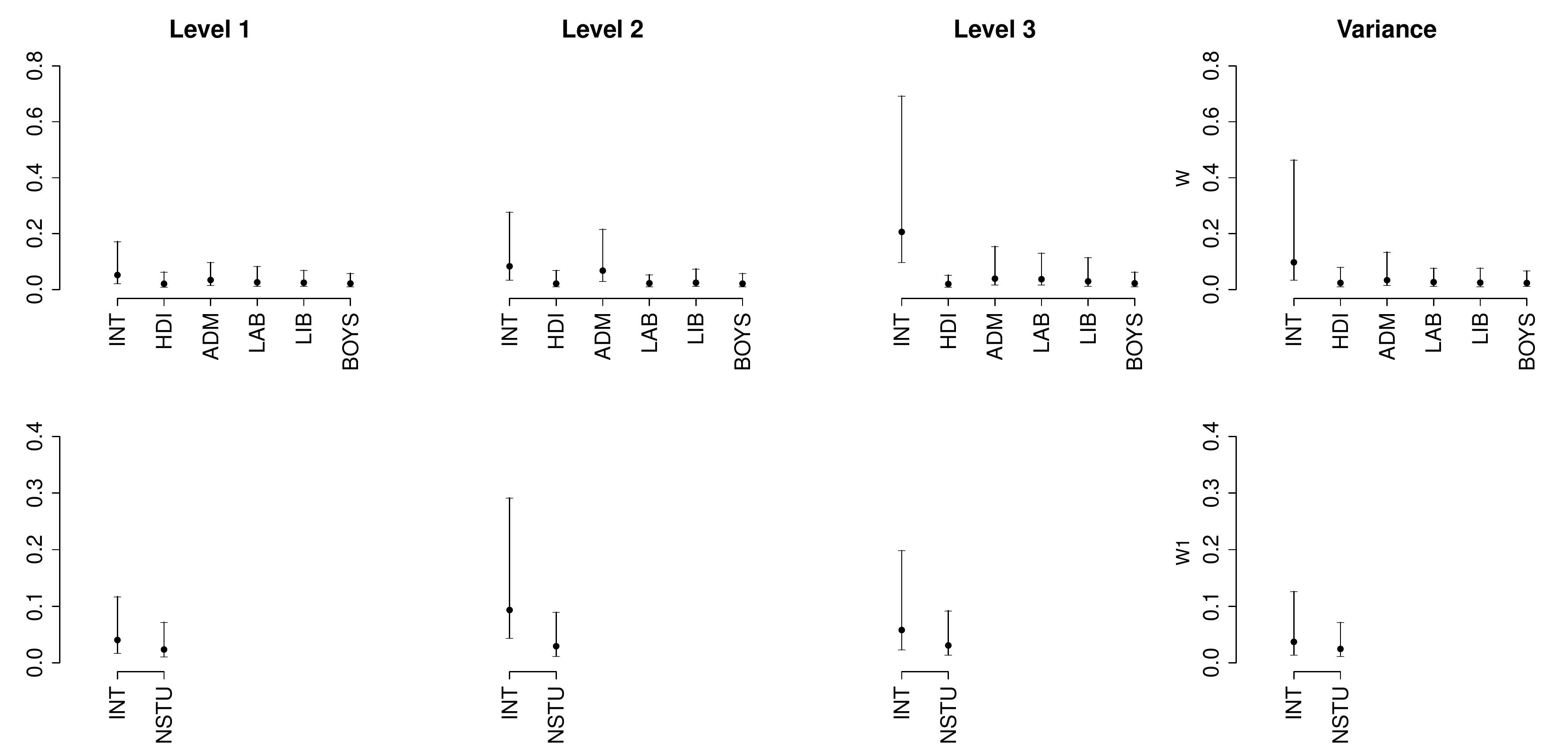}
\caption{Posterior summary (mean: solid circle, and limits of the 95\% credible intervals: solid lines) of the variances ($W_{\alpha m}$ and $W_{\gamma k}$, $m=0,1,\cdots,5$, and $k=0,1$) of the coefficients of the covariates in the logit of the mean (first row) and log of the precision (second row) parameters. See equations (\ref{eq:beta}),  (\ref{eq:evolution}), (\ref{eq:delta}), and (\ref{eq:evolutionprec}).\label{fig:hiper} }
\end{center}
\end{figure}

Panels of Figure \ref{fig:hiper} show the posterior summary of the variances of the coefficients, and the respective variances of the evolution equation, $W_{\alpha m}$ and $W_{\gamma k}$ ($m=0,1,\cdots,5$, and $k=0,1$), for  the coefficients in the logit of the  mean (first row), and in the log of the precision parameters (second row) (see equations (\ref{eq:beta}),  (\ref{eq:evolution}), (\ref{eq:delta}), and (\ref{eq:evolutionprec})). For both cases, the respective intercepts result in the highest values of the variances. 

\begin{figure}[!hbt]
\begin{center}
\includegraphics[scale=0.45]{\dir 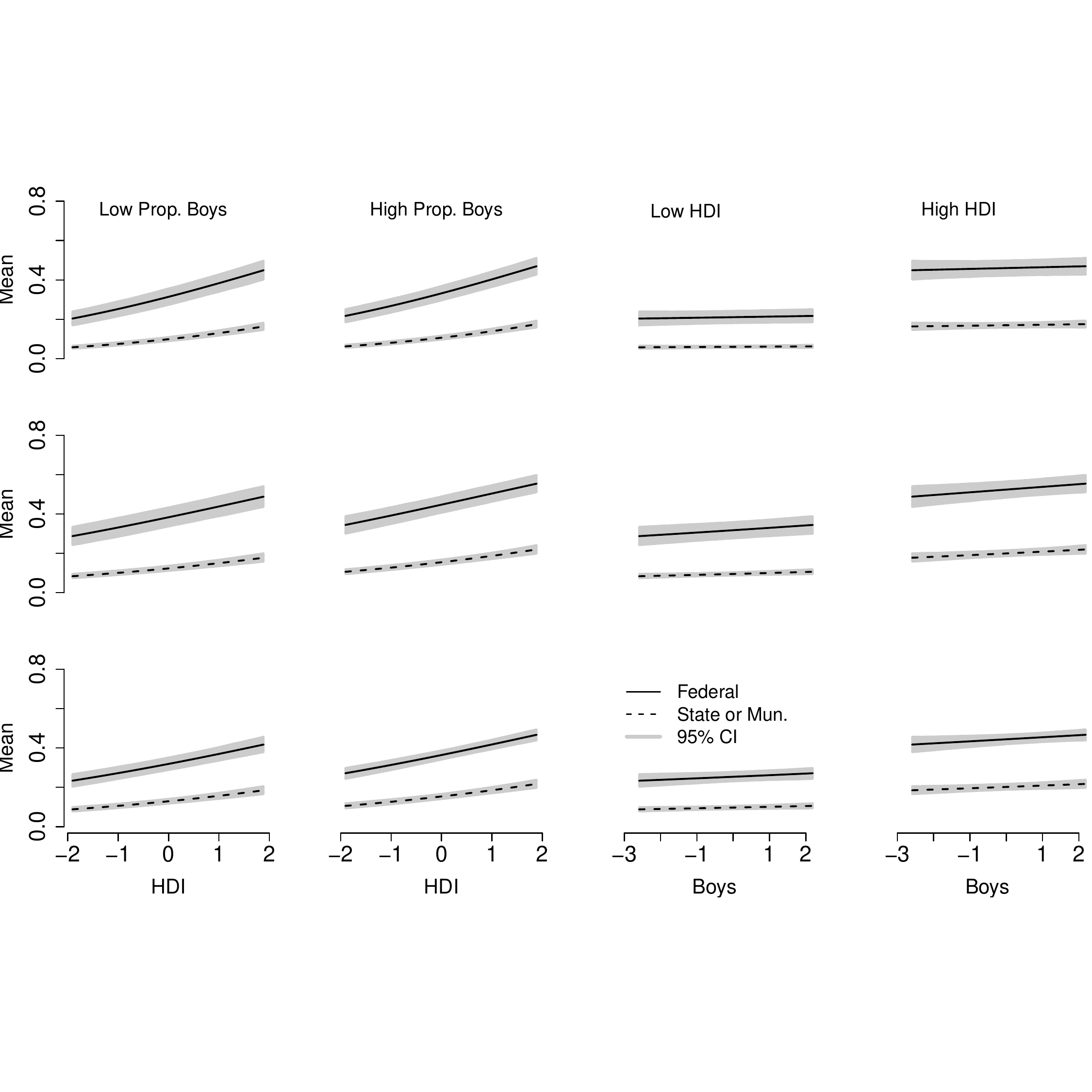}
\caption{Posterior summary (mean: solid (federal schools) and dashed (state or municipal schools) lines, and respective limits of the 95\% credible
intervals: shaded areas) of the mean, $\frac{\exp\left(\sum_{i=0}^5 {\bfbeta}_{it}'{\bf x}_{ijt}\right)}{\left(1+\exp(\sum_{i=0}^5 {\bfbeta}_{it}'{\bf x}_{ijt})\right)}$, for the different educational levels 1, 2, and 3 (rows),  year $t=8$ (year 2013), and different values of ${\bf x}_{ijt}'$=(1,ADM=1 or 0,HDI,Lab=1,Lib=1,Boys).\label{fig:FederalHDIEffect} }
\end{center}
\end{figure}

Panels of Figure \ref{fig:FederalHDIEffect} show the posterior summary of the mean, $\frac{\exp\left(\sum_{i=0}^5 {\bfbeta}_{it}'{\bf x}_{ijt}\right)}{\left(1+\exp(\sum_{i=0}^5 {\bfbeta}_{it}'{\bf x}_{ijt})\right)}$, under some particular scenarios,  for the different educational levels and last observed year, $t=8$ (year 2013). Panels in the first two columns of Figure \ref{fig:FederalHDIEffect} show the behaviour of the mean as a function of different values of the (standardized) HDI, considering  low and high values of the standardized proportion of boys in the second phase, for federal,  state or municipal schools,  and those with laboratory (LAB=1), and library (LIB=1). Clearly, regardless of the educational level (different rows) considered, federal schools show a steeper increase on the mean as the HDI increases, when compared to state or municipal ones. This leads to a greater difference in mean performance between federal and state or municipal schools located in municipalities with high values of the HDI.

Panels in the last two columns of Figure \ref{fig:FederalHDIEffect} show the behaviour of the mean as a function of different values of the (standardized) proportion of boys, for  low and high values of the standardized HDI, federal,  state or municipal schools,  with  laboratory (LAB=1), and library (LIB=1). In the first educational level (first row and columns 3 and 4) as the proportion of boys increases the mean tends to be constant, independent of the value of the HDI. Again, the effect of the HDI is very clear: regardless of the administrative level, schools located in municipalities with high values of HDI perform better than those in locations with low values of the HDI. Also, it is clear that there is a greater difference in mean performance between federal and state or municipal schools in locations with high value of the HDI, when compared to those with low value of the HDI. For the second and third educational levels (second and third rows and columns 3 and 4) it can be noticed a smooth increase of the mean as the proportion of boys increases. 

\begin{figure}[!hbt]
\begin{center}
\includegraphics[scale=0.45]{\dir 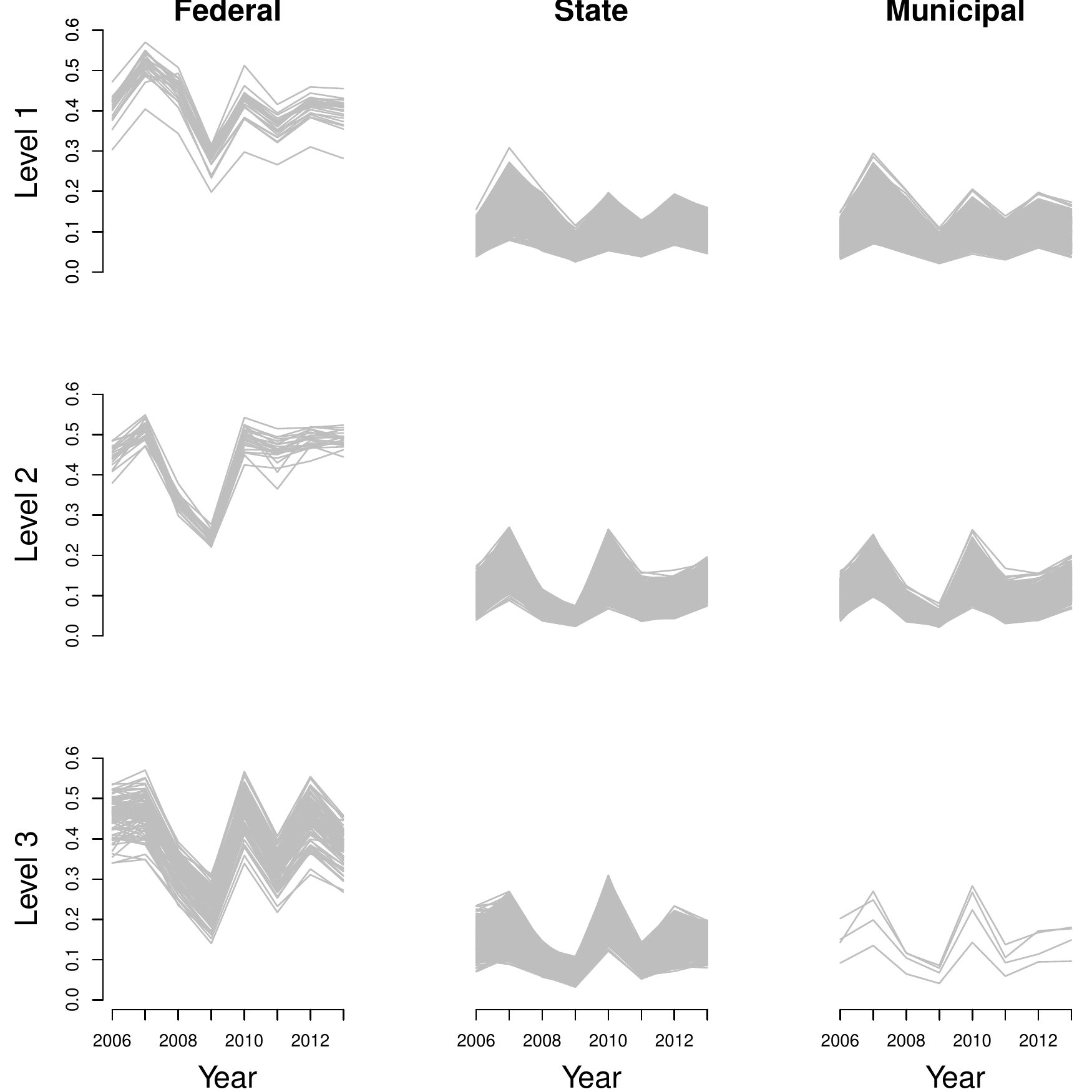}
\caption{Posterior mean of the mean score, $\mu_{ijt}$, for each school in the sample, per year, grouped by  educational (rows) and  administrative (columns) levels.\label{fig:schoolmeans} }
\end{center}
\end{figure}

Our model allows to compare the performance of schools according to their administrative and educational levels. Panels of Figure \ref{fig:schoolmeans} show the posterior mean of $\mu_{ijt}$ of each school in the sample, grouped by its educational and administrative levels. Clearly, independent of the educational level considered, federal schools have mean performance greater than state and municipal ones. However, for level 3, the difference in performance between federal and state schools seem to be smaller than this difference for levels 1 and 2. Also, the posterior mean of  federal schools in level 3 (first column and third row) show more variability than in levels 1 and 2. State and municipal schools have very similar performance, with scores varying below 0.3 for most of the editions. 
It is noticeable that all schools presented a drop in performance in the editions of 2008 and 2009.

\section{Discussion \label{sec:conclusion}}

This paper analyses the performance of schools that took part in eight editions of the second phase of the OBMEP, from 2006 until 2013. As different exams are given in different years, we propose an {\em ad hoc} standardization, per level and year, of the schools' mean scores such that they lie in the interval $(0,1)$.  It would have been better if organizers of the OBMEP used some tool to standardize the level of difficulty of the exams across years. This is an issue that should be tackled in the next editions of the OBMEP.

A hierarchical dynamic beta regression model is proposed to investigate the importance of some covariates in explaining the performance of schools in different educational levels. We allow the coefficients of the covariates to vary per level and year. We also explore models that allow the precision parameter of the beta distribution to be a function of the number of students in the school present in the second phase of the OBMEP.  Inference procedure is performed under the Bayesian paradigm and uncertainty about parameters' estimation is obtained in a straightforward fashion. A sample from the posterior distribution of the parameters was obtained through MCMC. In particular we used the Gibbs sampler with some steps of the Metropolis-Hastings algorithms. Implementation of the MCMC algorithm was done using the software {\tt JAGS}. Model comparison criteria suggest model M5, which assumes the logarithm of the precision parameter as a function of the number of students in the second phase of the OBMEP, with effects varying per level and year, perform best when compared to simpler versions of the proposed model.

Important conclusions are drawn from this study.  Overall, the mean performance of schools tend to be low, with scores varying below 0.3 for state and municipal schools in all three educational levels.  Results show that, in general, federal schools perform better than state or municipal ones. However, the posterior mean of federal schools in level 3 show more variability than in levels 1 and 2 (Figure \ref{fig:schoolmeans}). The difference between federal, and state and municipal schools tend to be greater in the first and second levels of the OBMEP. Federal schools show a steeper increase in their mean performance, as a function of the HDI, than state or municipal ones. As in Brazil the HDI is highly correlated with the geographical region, federal schools in the south and south-east (with higher values of HDI) regions of the country tend to perform better than federal schools in the north or north-east (with lower values of HDI) regions of Brazil. Also, state or municipal schools in locations with higher values of HDI tend to have mean performance  closer to those of  federal schools in locations with lower values of HDI, regardless of the proportion of boys the school has  in the second phase of the OBMEP (first two columns of Figure \ref{fig:FederalHDIEffect}). 

Schools with greater proportion of boys in the second phase of the OBMEP tend to perform slightly better in the second and third levels, with this covariate having no effect on  the mean of the scores for the first level (last two columns of Figure \ref{fig:FederalHDIEffect}).  
The possible difference in performance of boys and girls in mathematics exams has been the object of interest in different studies \citep{Hyde:Mertz:2009}. The analysis of the results of PISA 2009 show that boys outperformed girls in mathematics in 35 out of the 65 countries and economies that took part in PISA 2009. On the other hand, for 25 countries no significant difference was observed between the genders, whereas for 5 countries  girls outperformed boys in the  mathematics exam of PISA 2009 \citep{OECDGirls:2011}. 
Why the proportion of boys in the school affects positively the average scores in levels 2 and 3 of the OBMEP, when the students are slightly older and with a better understanding about their interests, clearly needs deeper investigation and understanding.




Our current interest is to investigate what kind of impact the OBMEP has on the Brazilian educational system. \cite{Biondi:Vasconcellos:Menezes:2012} quantify the effects of the 2007 edition of the OBMEP on the average math scores of the ninth-graders participating in {\em Prova Brasil}, which is a national exam applied by {\em INEP} to all Brazilian students in the $8^{th}$ and $9^{th}$ grades of publich schools. We plan to focus on students in the last year of high school. Considering different years we plan to use causal inference and propensity score methods \citep{hirano:imbens:2004} to investigate the effect of the OBMEP on the performance of students in different editions of the High School Brazilian National Exam ({\em Exame Nacional do Ensino M\'edio}, ENEM). Every year, results of the ENEM are used by nearly 500 universities in Brazil as a selection criterion for admission to higher education.






\section*{Acknowledgements}

This work is part of Moraes' M.Sc. Dissertartion under the supervision of  A.M. Schmidt and H. S. Migon. We are grateful for financial support from CNPq and FAPERJ (Schmidt), CAPES (scholarship to Moraes) and CNPq (Migon). The authors thank Professor Claudio Landim (IMPA, Brazil), Monica Souza (OBMEP) and Luiz L. R. da Concei\c c\~ao (OBMEP) for introducing  the problem and making the dataset available.

\bibliographystyle{rss}
\bibliography{References}

\end{document}